\documentclass[11pt,a4paper]{article}
\usepackage[hyperref]{acl2019}
\usepackage{times}
\usepackage{latexsym}
\usepackage{url}

\usepackage{amsmath}
\usepackage{graphicx}

\aclfinalcopy 

\title{Deep learning languages: a key fundamental shift\\ from probabilities to weights?}

\author{François Coste \\
  Univ Rennes, Inria, CNRS, IRISA / F-35000 Rennes \\
  \texttt{francois.coste@inria.fr} \\}

\date{}

\begin{document}
\maketitle
\begin{abstract}
Recent successes in language modeling, notably with deep learning methods, coincide with a shift from probabilistic to weighted representations. We raise here the question of the importance of this evolution, in the light of the practical limitations of a classical and simple probabilistic modeling approach for the classification of protein sequences and in relation to the need for principled methods to learn non-probabilistic models.
\end{abstract}

\bigskip

\section{Introduction}

Probabilistic models have been extremely important for successful language processing, from the first accomplishments in natural language processing (NLP), with $n$-grams and their successors \cite[chapters 22 and 23]{RusselNorvigAIModernApproach3rd}, to the tools routinely used today for the annotation of biological sequences in bioinformatics, based on the profile hidden Markov models (pHMM) or covariance models (CM) \citep{durbin_eddy_krogh_mitchison_1998,Coste2016}. Yet, after its success in computer vision and pattern recognition, deep learning is now replacing probabilistic models to become the new key player in the NLP field, showing that neural networks based on dense vector representations are very powerful for performing a large variety of NLP tasks. Impact of deep learning is until now limited in biology, but promising advances are being made (see for instance  \citet{doi:10.1098/rsif.2017.0387}, notably section 3 for the applications to biological sequences) and we can expect progress in NLP to be beneficial for biological sequence analysis once again.\\

While the main appealing feature of neural networks was initially \emph{non-linearity}, the historical ``S''-shape activation functions (such as the $\tanh$ function) are more and more replaced by the Rectified Linear Unit (ReLU) function, reducing non-linearity to its simplest form: a function returning $0$ or a weighted linear combination of its input values. The most important feature of deep learning in NLP seems now to be its ability to learn intermediate representations ---such as the word vector representations (word embeddings) learned by word2vec \citep{word2vec} and GloVe \citep{pennington2014glove}, or the recent deeper contextual representations learned by ELMo \citep{Peters:2018}, ULMFiT \citep{howard2018universal}, and GPT \citep{gpt,gpt2}--- in a hierarchical manner up to the last layers of the neural networks in charge of their weighted combination for solving the task at hand. Using weighted combination of weight vectors (or matrices) optimized by gradient descent appears to be more and more the core of deep learning, as witnessed for instance by the growing importance of tensor-based libraries and of automatic differentiation \citep{DBLP:journals/jmlr/BaydinPRS17}.

Hierarchical representations are common in formal grammars and gradient descent is not an original approach. The key advantage of deep learning with respect to the inference of classical grammatical representations could come from using and combining \emph{weights} rather than \emph{probabilities}. We think that better understanding and evaluating the importance of this difference is essential for future research in this area. As a first contribution, we present here practical problems illustrating the fundamental limitations of current probabilistic modeling approaches for language learning. The question we raise is whether they could be properly handled with more elaborate probabilistic modeling, or whether they demonstrate the interest of the deep "Probxit" towards weighted models.

\section{Modeling with probabilistic grammars}
To cope with inherent "noise", "variations" and "errors", languages have long been modelled with probabilistic models. 
In practice, these models rarely exceed the expressiveness of probabilistic context-free grammars, or even probabilistic automata. We introduce briefly the main related definitions and notations following \citet{Vidal:2005:PFM:1070615.1070803,Vidal:2005:PFM:1070615.1070798} to which we refer the reader for a more detailed presentation of the different probabilistic grammatical models of languages.

A probabilistic context-free grammar (PCFG) $G$ is defined as a tuple $\langle Q, \Sigma, S, R, P \rangle$ where $Q$ is a set of non-terminal symbols, $\Sigma$ is a finite alphabet, $S \in Q$ is the initial symbol, $R$ is a set of rules $A \rightarrow \alpha$ with $\alpha \in (Q \cup \Sigma)^*$ and $P : R \mapsto [0,1]$ define the probabilities of rules for each $A\in Q$, under the constraint:
\begin{equation}\label{eq:derivprobdistrib}
  \forall A \in Q \colon \sum_{(A\rightarrow\alpha) \in R} P(A\rightarrow\alpha) = 1.
\end{equation}
A successful (leftmost) derivation of a sequence $x$ from $S$ by a succession of rules $r_1\ldots r_l \in R^l$ is denoted by $S \overset{r_1\ldots r_l}{\Rightarrow} x$ and its probability is $P(r_1\ldots r_l)=\prod_{i=1}^{l} P(r_i)$. If a sequence $x$ can be successfully derived from $S$, its probability is $P(x)=\sum_{r\in R^+\colon S\overset{r}{\Rightarrow} x}P(r)$, otherwise its probability is $0$. Only consistent PCFGs, i.e.\ satisfying:
\begin{equation}\label{eq:seqprobdistrib}
  \sum_{x\in\Sigma^*}P(x)=1
\end{equation}
defining thus a probability distribution on $\Sigma^*$, are classically considered. 

Probabilistic regular grammars, more often depicted as probabilistic automata (PA), are PCFGs with simpler rules of the form $A\rightarrow aB$, $A\rightarrow a$, or $A\rightarrow \lambda$  with $A,B \in Q$, $a\in \Sigma$ and $\lambda$ denoting the empty sequence.

\begin{figure*}[t]
  \centering
  \includegraphics[width=0.93\textwidth]{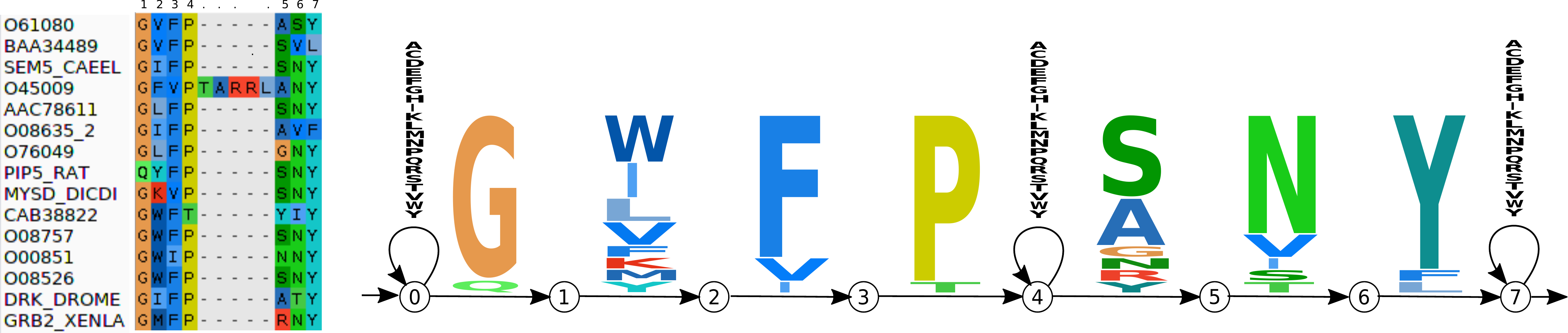}
  \caption{pHMM-like probabilistic automata (right) from a toy alignment of 15 SH3 sequences with 7 conserved columns (left). Height of amino-acid letters is proportional to the probability of the transition labeled by the letter. Here, probability of amino-acids in left-to-right transitions is their observed frequency in the modelled column while they are assumed equiprobable in the self-loop transitions introduced at beginning and end of the PA, to parse complete proteins, and in its middle, to allow insertions between columns 4 and 5 (as in protein O45009).}
  \label{fig:SH3aliPA}
\end{figure*}
To give intuition, we will illustrate some limitations of purely probabilistic approaches on a practical task: modeling a family of protein sequences. This task is classically achieved with pHMMs, which are discrete HMMs with a predefined left-to-right topology. Illustration will focus on the core of this approach, simplified and transposed to PA to remain in the grammatical formalism. As an example of a family of protein sequences, we consider the SH3 domain, a region fairly well conserved among several protein sequences that is known to be important for binding and interacting with other proteins. This example has been realistically covered in the pHMM tutorial by \citet{krogh1998introduction}. We will simply use here a toy ``profile PA'' built from an excerpt of the alignment of the SH3 regions from some protein sequences, shown in figure \ref{fig:SH3aliPA}. The choice of this example was obviously driven by our own domain of expertise, but also because it enables us to illustrate modeling problems with respect to ''objective'' physico-chemical considerations. The problems presented in the following sections should still be generic enough to be easily transposed to other applications and methods using probabilistic sequence models.









\section{Limitations}

We present here a list of practical limitations of classical probabilistic language modeling that should not arise when using weighted models. This does not mean that more elaborate probabilistic approaches could not solve these issues, nor that the best way to learn weighted models overcoming these limitations is known. We see these problems as critical testbeds for studying and comparing both approaches in order to better measure the contribution of the shift from probabilities to weights in the success of deep learning.
Ideally, we would like these examples to also help developing principled approaches for learning weighted grammatical models, such as those available for probabilistic models. A pragmatic first step in this direction could be the elaboration of well-founded schemes for parameter estimation of ``profile weighted automata'' outperforming the simple maximum likelihood estimation of pHMMs' probabilities (or, better, the sophisticated Bayesian inference methods used to boost their performances by the addition of pertinent pseudocounts from Dirichlet mixture priors, as initiated for instance by \cite{SjolanderKBHKMH96}).

\paragraph{L1: Probability scattering at each non-deterministic derivation step}

Comparing the probabilities given by probabilistic grammars to sequences of different lengths is a well-know practical problem. The profile PA in figure \ref{fig:SH3aliPA} will assign smaller probabilities to longer proteins even if they have exactly the same SH3 region, because of the loss of probability in self-loop transitions consuming the rest of the protein sequence. This problem is tackled in bioinformatics by looking at the ratio between the probability of the sequence and its probability according to a null model: these ratios are usually directly integrated in the profiles by associating log-odds scores to transitions rather than probabilities. Profile are thus already used under the form of weighted models which combine two probabilistic models.

More generally for PCFGs, equation \ref{eq:derivprobdistrib} implies a probability loss by a factor strictly smaller than one, at each derivation step involving a non-deterministic rule. Such derivations scatter more and more probabilities between alternative suffix languages, resulting in negligible probabilities for longer derivations. Moreover, they induce a dependence of probability values to the number of non-deterministic derivations, which has often to be corrected or compensated in practical tasks, while this could be avoided by not satisfying equation \ref{eq:derivprobdistrib}.



\paragraph{L2: Dependence to number of choices for mass probability distribution within alternative rules}
Another consequence of equation \ref{eq:derivprobdistrib} is that the probability mass has to be shared between all alternative rules. Its amount by rule depends then from the number of alternatives. For the characterization of SH3 regions, figure \ref{fig:SH3aliPA} shows the importance of having the amino-acid {\tt G} at the first position. But when we look at the second conserved column, we can see that it contains only hydrophobic amino-acids and it might be as important for SH3 function to have a hydrophobic residue at the second position as a {\tt G} at the first position. This cannot be modeled with PCFGs using the amino-acid alphabet: assuming that there are 14 hydrophobic acids and that they are equiprobable, the probability of transition with these amino-acids will be at most $\frac{1}{14}$ making a small difference with the completely equiprobable $\frac{1}{20}$ null model, in contrast to $\frac{14}{15}$, or any value close to $1$, that can be assigned at the first position to a transition with {\tt G}. We could imagine working on a larger alphabets, such as the amino-acids powerset, but the exponential growth of the alphabet size would then complicate training. Considering weights (e.g. reflecting adequacy of particular amino-acids at position, instead of occurrence) could be a more efficient solution in practice. 



\paragraph{L3: Identical probability mass for all choice points}
Because of the normalisation to $1$ in equation \ref{eq:derivprobdistrib}, all choice points have the same total weight in final probability. Yet, some choices can be more important than others. For instance, prior expert knowledge could tell us that amino acid {\tt P} in the fourth position is more important for the function of SH3 than amino-acid {\tt G} in the first position, but this could not be modelled with a PCFG. Authorizing a normalisation to variable values (e.g. reflecting relevance of position modelled with a particular non-terminal) would help here. Note that it could also be an indirect way to overcome limitation L2.


\section{Towards language modeling Probxit?}

\paragraph{Analysis of limitations}
Limitations seen above are induced by equation \ref{eq:derivprobdistrib} which enables proper PCFGs to satisfy equation \ref{eq:seqprobdistrib} and thus define a probability distribution over all sequences. While discarding the constraint of equation \ref{eq:derivprobdistrib} is tempting, this constraint is intrinsically needed to distribute the remaining probability mass between all sequences that can be generated after each choice point, and it makes perfect sense to model the probability of sequence occurrence in the language with respect to the uncertainty brought by the non-deterministic rules.

Actually, what is illustrated here is more a misused feature of probabilistic grammars rather than a real limitation of these models: probability of sequences under constraint of equation  \ref{eq:derivprobdistrib} specifies the likelihood of their occurrence inside the language, not their probability of being in the language. Considering the SH3 example, probabilities tell us how often we can expect to see a given SH3 sequence comparatively to other SH3 sequences, not to estimate the uncertainty that the sequence can or not perform the SH3 function. If we are interested in membership to language, it is clearly given by the non-probabilistic part of the grammar: a sequence $w$ is in the language $L$ if it can be successfully derived by rules which have non-zero probability. If we were interested in the uncertainty of membership, we should consider learning a probability $P(x \in L)$, with $P(x \in L) + P(x \not\in L) = 1$ instead of equation \ref{eq:seqprobdistrib}. 
Likelihood of occurrence in the language can thus be considered at most as an imperfect proxy for membership prediction. 

\paragraph{Research directions}
The first research direction to overcome the enumerated limitations is to work on learning the topology of the grammars. Note that this could be completed by learning also a membership probability, but the meaning of this uncertainty, and how it could be learned, would have to be clarified. It could be a measure of the uncertainty of decision with respect to current knowledge (e.g.\ based on the number of available examples supporting the decision). In the example of SH3, we can also imagine, for instance, that the binding affinity of different SH3 regions could influence their probability of binding to another protein and to perform their function: this could be formalized as a kind of membership probability, but would require information about the probability of performing the function in the training set. We will not consider these extensions here and will continue to focus on membership problem.

Learning a good grammar topology can be hard. In contrast, adding probabilities (and a recognition threshold) to simpler topologies has been shown successful to improve sufficiently their expressiveness for many classification tasks
, while maintaining the number of parameters low enough to be estimable from training sets. This strategy has been shown efficient in practice, despite the limitations that we have identified. A second research direction is to study if more elaborate probabilistic approaches could be used more successfully here. The third research direction is to get rid of the constraint of equation \ref{eq:derivprobdistrib}, which is not really needed for membership prediction, to learn this way weights instead of probabilities of grammar rules.

In this third approach, the weights of sequences could be divided by a partition function (if it exists) to be considered as probabilities taking values in the interval $[0,1]$ and satisfy equation \ref{eq:seqprobdistrib}. Besides the problem of the existence and the computation cost of such a partition function, we think that the constraint of equation \ref{eq:seqprobdistrib} has also to be discarded, since it restricts the expressiveness of the approach. Indeed, (positive) weights are useful if a threshold greater than zero is chosen so as sequences accepted in the language are those whose weight is above this threshold (otherwise, successful derivation by transitions with non-zero weight is sufficient). By transforming weights into probabilities, the choice of a threshold $t$ would then only enable to define finite languages (with at most $\frac{1}{t}$ sequences since total probability mass is $1$), restricting thus unnecessarily the languages that can be represented.

\paragraph{From probabilities to weights}
Getting rid of both equations means shifting from probabilistic to weighted representations. This move is visible in research on learning weighted grammars (see for instance contrastive estimation from \citet{Smith:2005:CET:1219840.1219884}, still motivated by mass distribution constraint) and can be considered an intrinsic feature of deep learning approaches. While practical successes show the interest of these approaches, better understanding their fundamental contribution and being able to develop principled approaches for learning weighted grammars, such as those developed for probabilistic grammars, remains a challenge. A key we identified is not (mis)using occurrence probabilities anymore and we exhibited some advantages of using weighted rather than probabilistic models, but theoretical and practical studies are still needed. The problem of training weighted, instead of probabilistic, profile automata on proteins seems a good support for these studies since it is simple and enables objective evaluation of the approaches. To address these questions, we think that we will need first to find the equivalent for weighted grammars of maximum likelihood or Bayesian approaches (especially for the incorporation of prior knowledge, an essential feature of using pHMMs on proteins) and the nice compositional properties of probabilistic approaches, and would really like to discuss this during the workshop.
\clearpage
\section*{Acknowledgments}
I am grateful to Hugo Talibart and Witold Dyrka for fruitful discussions that motivated this work submitted as a four pages opinion paper to ACL 2019 Workshop ``Deep Learning and Formal Languages: Building Bridges''.
\bibliography{biblio}
\bibliographystyle{acl_natbib}
\end{document}